\documentclass[12pt]{article}
\usepackage[top=1in, bottom=1in, left=1in, right=1in]{geometry}                
\usepackage{amssymb}
\usepackage{url}
\usepackage{longtable}
\usepackage{ragged2e}
\usepackage{tabularx}
\usepackage{array}
\usepackage{supertabular}
\usepackage{amsmath}
\usepackage{bm}
\usepackage{enumerate}
\usepackage{multirow, threeparttable}
\usepackage{amsmath}
\usepackage{amssymb}
\usepackage{graphicx}
\usepackage{textcomp, setspace}
\usepackage{float}
\usepackage{pgf,tikz}
\usepackage{amsthm}
\usepackage{listings}
\usepackage{natbib}

\newtheorem{procedure}{Procedure}
\newtheorem{algorithm}{Algorithm}

\DeclareMathOperator*{\argmin}{argmin}

\title{Subtask Analysis of Process Data Through a Predictive Model}
\author{Zhi Wang, Xueying Tang, Jingchen Liu and Zhiliang Ying}
\date{}

\begin{document}
	\maketitle
\begin{abstract}

Response process data collected from human-computer interactive items contain rich information about respondents' behavioral patterns and cognitive processes. Their irregular formats as well as their large sizes make standard statistical tools difficult to apply. This paper develops a computationally efficient method for exploratory analysis of such process data. The new approach segments a lengthy individual process into a sequence of short subprocesses to achieve complexity reduction, easy clustering and meaningful interpretation. Each subprocess is considered a subtask.
The segmentation is based on sequential action predictability using a parsimonious predictive model combined with the Shannon entropy. Simulation studies are conducted to assess performance of the new methods. We use the process data from PIAAC 2012 to demonstrate how exploratory analysis of process data can be done with the new approach. 

\end{abstract}
	
\section{Introduction}
	Technology advances in educational assessments expand measurable skills beyond conventional ones. For instance, 14 items for Problem Solving in Technology-Riched Environment (PSTRE) are included in the 2012 Programme for the International Assessment of Adult Competencies (PIAAC). In these items, test-takers complete real-life tasks in various simulated interactive environments. 
	These interactive items not only facilitate test administration but also enrich data available for analyses. Once a test-taker finishes an item, we observe not only whether or not the tasks are completed successfully but also the entire problem-solving processes recorded in log files. 
	
\begin{figure}[htb]
	\includegraphics[width=\textwidth]{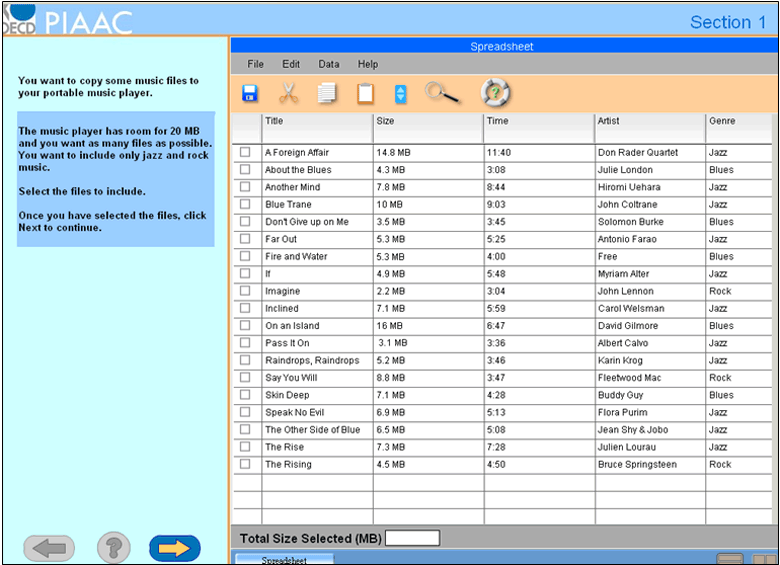}
	\caption{Screenshot of a sample item in PIAAC 2012. Source: \protect\url{https://nces.ed.gov/surveys/piaac/sample_pstre.asp}}\label{fig:sample_item}
\end{figure}

	Figure \ref{fig:sample_item} is a screenshot of the interface of a released item in PIAAC. The item description in the left panel instructs respondents to select music files from the spreadsheet in the right panel for copying to a music player with 20 MB storage. The respondents are required to select only jazz and rock music with maximum number of files. 
To complete the task, one can sort the files in an ascending order of the file sizes by clicking the sort button in the toolbar and then select the first several files of genre jazz or rock until the total sizes of selected files (indicated at the bottom of the interface) reach 20 MB. 
Once finished, the respondent can move to the next item by clicking the right arrow button at the bottom of the left panel.
The response process is recorded in a log file as a sequence of actions. Under the above-mentioned strategy, the action sequence is ``Start, Toolbar\_Sort, Sort\_A\_2, Sort\_OK, Menu\_Edit, Menu\_Help, Menu\_Data, Toolbar\_Help, Toolbar\_Save, Tick\_9, Tick\_12, Tick\_19, Tick\_13, Tick\_14, Next''. The subprocess ``Toolbar\_Sort, Sort\_A\_2, Sort\_OK'' is related to sorting the files. In particular, action ``Sort\_A\_2" means to sort the music files in an ascending order of the values in the second column. After sorting, the respondent learns to select files and performs ``Menu\_Edit, Menu\_Help, Menu\_Data, Toolbar\_Help, Toolbar\_Save" to explore the interface. The final subprocess ``Tick\_9, Tick\_12, Tick\_19, Tick\_13, Tick\_14'' is related to selecting files. This action sequence details a respondent's problem-solving process and constitutes one observation of the response processes. 

Process data in educational assessment have gained great prominence as they provide a new venue for investigating and understanding problem-solving behaviors.
Tools in natural language processing such as $n$-grams have been adapted to distinguishing responses process belonging to different groups \citep{he2016analyzing,stadler2019taking,liao2019mapping}. Automated feature extraction methods have been proposed to facilitate the analysis of response processes through regression models  \citep{tang2019mds,tang2020ae}. \citet{qiao2018data} applied various machine learning techniques to process data to investigate strategies used by students within the same category or across different categories. \citet{chen2019statistical} proposed an event history model for predicting response duration and outcome. 
A common goal of these works is to study the relationship between response processes and other variables of interest such as response outcomes, response time, and respondents' demographics. 

Responses processes contain a great amount of variation and noise. This is because simulated computer environment often allows a great deal of flexibility. 
The tasks for PSTRE items are often complicated and require several necessary steps or \emph{subtasks} to accomplish. 
It is natural to decompose the total variation among response processes into within subtask variation and between subtask variation, borrowing the main idea from the one-way analysis of variance.
In the PIAAC example of Figure 1, three subtasks are identified: SORT, EXPLORE and SELECT. Their interpretation and the corresponding subprocesses are summarized in Table \ref{table:subtask_sample}.
\begin{table}[h]
	\centering
	\caption{Subtasks of the sample item.}\label{table:subtask_sample}
	\begin{tabular}{llp{0.43\linewidth}}
		\hline
		Subtask & Description & Subprocess example\\
		\hline
		SORT & Rearrange spreadsheet by sorting. & Start, Toolbar\_Sort, Sort\_A\_2, Sort\_OK \\
		EXPLORE & Get familiar with interface. & Menu\_Edit, Menu\_Help, Menu\_Data, Toolbar\_Help, Toolbar\_Save\\
		SELECT & Select files by ticking boxes. & Tick\_9, Tick\_12, Tick\_19, Tick\_13, Tick\_14, Next\\
		\hline
	\end{tabular}
\end{table}

Identifying subtasks in a response process helps us understand problem-solving strategies and explore its relationship with the test performance and other characteristics. 
However, the observed processes are not naturally segmented. Although subtasks and strategies can be identified by domain experts based on their understanding of the item design and human cognitive processes, 
such an approach is time-consuming and not scalable. Moreover, purely subjective identification often results in confusing and contradictive conclusions.

In this paper, we propose a data-driven procedure to decompose each response process into several subprocesses,  whereby identifying the corresponding subtasks.
The procedure is motivated by observing that actions within a subtask are more predictable than those at  the transition between two subtasks.
That is, if an action is highly predictable, then it is likely to belong to the same subtask as the preceding actions; if an action is not predictable, then it suggests a transition to a different subtask.
This predictability-based segmentation turns a long and noisy action sequence into a short and simple sequence of problem-solving subtasks that are easy to visualize and analyze. 
The order and frequency of subtasks reveal respondents' problem-solving strategies and facilitate the discovery of their association with problem-solving strategies and other characteristics.

The performance of the proposed procedure is examined and compared to a hidden Markov model (HMM) in both simulation and a case study of the PSTRE items in PIAAC 2012. The dataset contains the response processes from more than forty thousand respondents. The lengths of the response processes range from several dozen to a few hundred, depending on the task complexity and individuals' problem-solving strategies. We demonstrate in the case study how the results produced by the proposed methods could help understand problem-solving behaviors. We visualize the subtasks and analyze their association with a variety of variables such as response outcome and response time. The decomposition can further be used to guide potential item design and educational interventions. 

The rest of this article is organized as follows. 
Section \ref{sec:method} describes the subtask identification method. In Section \ref{sec:sim}, we evaluate the proposed method on simulated data. A case study of PIAAC 2012 is presented in Section \ref{sec:case_study}. Section 5 contains some concluding remarks.


\section{Subtask analysis}\label{sec:method}




In this section, we develop a data-driven method to identify subtasks. The underlying rationale is that the subtasks are usually simpler and their response processes are more homogeneous. From a technical viewpoint, short-term action predictions are more accurate within the same subtask than those at the transition between subtasks. Therefore, we begin with an action predictive model, followed by the predictability-based segmentation or subtask identification method.

	Through out this paper, let $\mathcal{A} = \{a_1, \ldots, a_M\}$ denote the collection of possible actions for  a problem-solving item, where $M$ is the total number of possible actions for this  item.
	We use $\bm s = (s_1, s_2, \ldots, s_T)$ to denote a generic response process, where $s_t$  is the $t$-th action in the process and takes its values in $\mathcal{A}$, and $T$ is the sequence length.
	We use $s_{1:t}$ as an abbreviation of $s_1, \ldots, s_t$ to denote the actions  up to time step $t$. 
	We observe $N$ response processes, denoted by $\mathcal{S} = \{\bm s^{(i)}, i = 1, \ldots, N\}$, where $\bm s^{(i)} = (s_1^{(i)}, \ldots, s^{(i)}_{T_i})$ is the  $i$-th observation. The sequence length varies among different observations. 

\subsection{Action prediction model}\label{sec:action_pred}
To begin with, we describe a predictive model for future actions that forms a basic building block of the process segmentation method. 
This means to specify, for each $t$,  the conditional distribution of $s_{t+1}$ given $s_{1:t}$, ${p(s_{t+1} \mid s_{1:t})}$. 
To obtain a parsimonious model, we first compress $s_{1:t}$ to a $K$-dimensional vector $\bm \theta_t$ and then use the multinomial logistic model \citep[MLM;][]{mccullagh2018generalized}, 
\begin{equation}\label{eq:logi}
p_{tj}  = P(s_{t+1} = a_j \mid s_{1:t})= \left\{
\begin{array}{ll}
\frac{\exp\left({{\boldsymbol{\beta}_j}^\top \boldsymbol{\theta}_t + \alpha_j }\right)}{1 + \sum_{i=1}^{M-1} \exp\left({{\boldsymbol{\beta}_i}^\top \boldsymbol{\theta}_t + \alpha_i }\right)}, & j = 1, \ldots, M-1; \\
\frac{1}{1 + \sum_{i=1}^{M-1} \exp\left({{\boldsymbol{\beta}_i}^\top \boldsymbol{\theta}_t + \alpha_i }\right)}, & j = M,
\end{array}
\right.
\end{equation}
where  $\alpha_j$ and $ \bm \beta_j$ are parameters. The construction of  $\bm \theta_t$, as described below, makes use of an embedding method \citep{bengio2003neural,kraft2016embedding} and a recurrent neural network \citep[RNN;][]{bengio1994learning}.

First, we associate each action $a_j$ with a $K$-dimensional vector $\bm e_j$, which can be viewed as a continuous representation (embedding) of the categorical $a_j$ 
and are estimated together with other parameters in the model. 
The response process $s_{1:T}$ is thereby transformed into a sequence of vectors, $\bm x_1, \ldots, \bm x_T$, where $\bm{x}_t$ is the embedding of $s_t$.
	Let $\bm X = (\bm x_1, \ldots, \bm x_T)^\top$ and $\bm E = (\bm e_1, \ldots, \bm e_M)^\top$. The embedding step, expressed by matrix multiplication, is $\bm X = \bm S \bm E$,  where $\bm S$ is a $T\times K$ binary matrix with $(t,j)$-th element being $1$ if and only if $s_t = a_j$.



Second, we recursively summarize the information in the embedding sequence $\bm x_1, \ldots, \bm x_t$ into $\bm \theta_t$ through an RNN. 
More specifically, $\bm \theta_t$ is obtained by synthesizing the information contained in the previous actions $\boldsymbol{\theta}_{t-1}$ with the current action embedding $\boldsymbol{x}_{t}$ through a function $\boldsymbol{f}$
\begin{equation}\label{eq:f}
\boldsymbol{\theta}_{t} = \boldsymbol{f} (\boldsymbol{\theta}_{t-1}, \boldsymbol{x}_{t}; \bm \gamma),
\end{equation}
where $\boldsymbol{\theta}_0$ is an initial value which may set as a vector of zeros and $\bm \gamma$ is a vector of parameters. The functional form of $\bm f$ is determined by a specific RNN structure. Here, we choose the gated recurrent unit \citep[GRU;][]{Cho2014phase} in our action prediction model; see Appendix \ref{sec:appendix:gru} for details. Note that $\bm \gamma$ does not depend on $t$; therefore, the number of parameters used to characterize the entire sequence $p(s_{t+1} \mid s_{1:t}), t \geq 1, $ does not depend on $T_{\text{m}}$. 
To summarize, we let $\mathcal{R}_t(\bm X; \bm \gamma)$ denote the resulting vector after applying \eqref{eq:f} $t$ times to the row vectors in $\bm X$ and write
\begin{equation*}
\bm \theta_t = \mathcal{R}_t(\bm S\bm E; \bm \gamma).
\end{equation*}
The parameters in the model described above contain the action embeddings $\bm E$, parameters in the RNN $\bm \gamma$, and the parameters in the multinomial logistic model $\alpha_j, \bm \beta_j$, $j = 1, \ldots, M-1$. Let $\bm\eta$ denote the vector that collects all parameters. Given a set of observed response processes $\mathcal{S} = \{\bm s^{(i)}, i = 1, \ldots, N\}$, the parameters are estimated by maximizing the log likelihood function
\begin{equation}\label{eq:generic loglik_peudo}
l(\bm\eta; \bm s^{(i)}) = \sum_{t=1}^{T_i-1}  \log p(s_{t+1} = s_{t+1}^{(i)} \mid s_{1:t} = s_{1:t}^{(i)}; \bm\eta) = \sum_{t=1}^{T_i - 1}\sum_{j = 1}^M \delta_j(s_{t+1}^{(i)}) \log p_{tj}^{(i)}.
\end{equation}
$\delta_j(s)$ is an indicator function for $s \in \mathcal{A}$
\begin{equation}\label{eq:indicator}
\delta_j(s) = \left\{
\begin{array}{ll}
1, & \text{if } s = a_j;\\
0, & \text{if } s \neq a_j,\
\end{array}
\right.
\end{equation}
$p_{tj}^{(i)}$ is obtained by replacing $\bm \theta_t$ with $\bm \theta_t^{(i)} = \mathcal{R}_t(\bm S^{(i)}\bm E)$ in \eqref{eq:logi} and $\bm S^{(i)}$ is the binary matrix representation of $\bm s^{(i)}$. See Appendix \ref{sec:appendix_estimation} for details.

\subsection{Sequence segmentation and subtask identification}\label{sec:seg}
The preceding model provides us, for each $t$, with a predictive (conditional) probability density $\bm p_t = (p_{t1}, \ldots, p_{tM})$.
We do not use this model to actually predict future actions but rather to assess the intrinsic uncertainty. To this end, we propose to use the Shannon entropy \citep{cover2006info}, 
\begin{equation}\label{eq:entropy}
h(\boldsymbol{p_t}) = - \sum_{j=1}^M p_{tj} \log p_{tj},
\end{equation}
which is used in information theory to quantify the uncertainty of a distribution.
For instance, a point mass distribution has entropy equal to  zero that is the minimum value, whereas
the maximum entropy is achieved at the uniform distribution.

\begin{figure}[t]
	\centering
	\includegraphics[trim={0cm 8cm 0cm 3cm}, width=0.95\textwidth, clip]{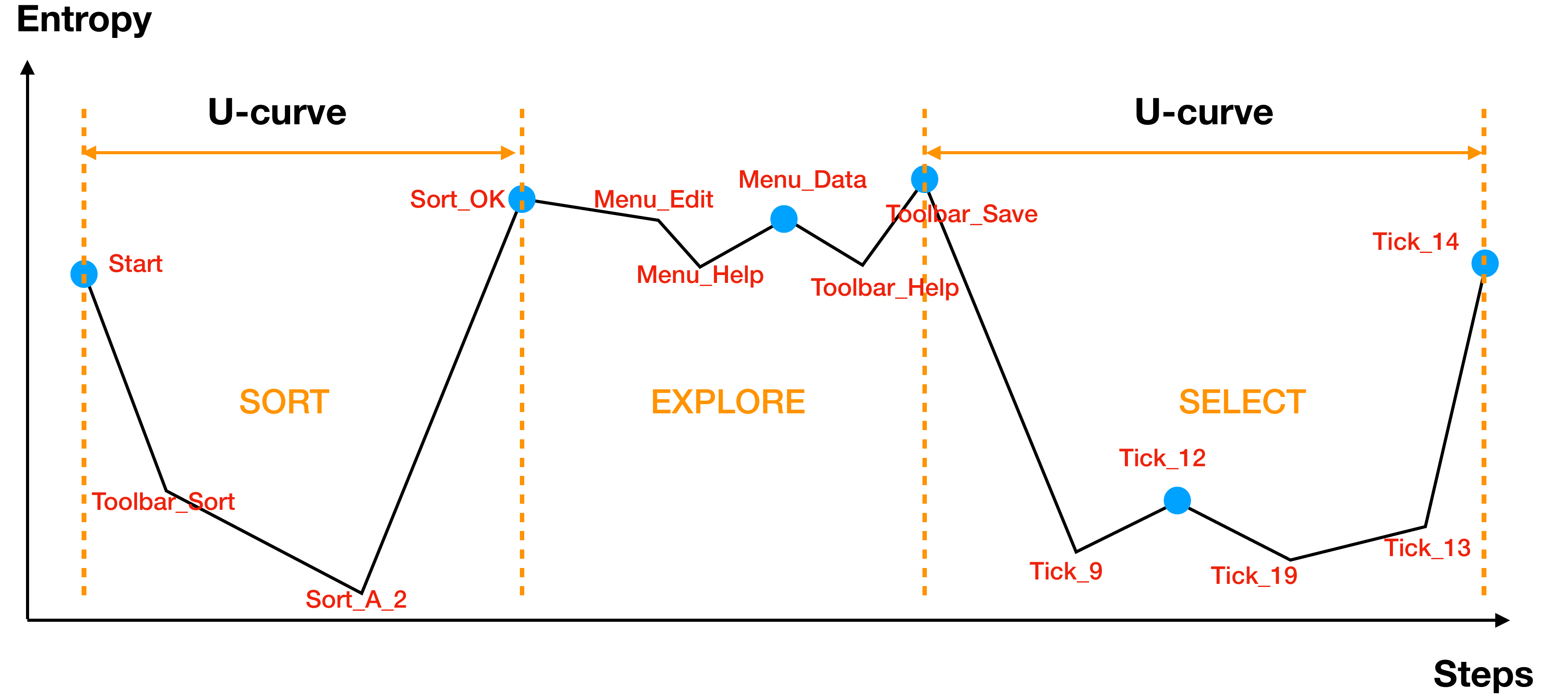}
	\caption{An example of a segmented entropy process. The blue dots represent the local maxima of the entropy process. The U-curves are identified by Algorithm \ref{al:seg} with $\lambda  = 0.5$.}
	\label{fig:U_curve}
\end{figure}

	Figure \ref{fig:U_curve} presents an evolution of entropy process $\bm h = (h_1, \ldots, h_{T-1})$ of a response process as a function of time $t$. 
	As the figure shows, a response process usually starts at a relatively high entropy since there is little information available the first action.
	Then the predictability of subsequent actions gradually increases and the entropy decreases accordingly as a subtask begins. 
	As the problem-solving process evolves, entropy rises to a relatively high level again suggesting that the person has accomplished a subtask and is about to explore the system for the next subtask. The entropy fluctuates several times until the process reaches an end.
	The entropy process $\bm h$ consists of several U-shaped curves, each of which corresponds to a subtask as marked in Figure \ref{fig:U_curve}.
	We observe that the processes are often very predictable within the same subtask. On the other hand, when a subtask is accomplished, it is generally more difficulty to predict the subsequent actions as there are usually several options and the test taker might take any of them at random. This forms our basic understanding of a response process. Each U-shaped curve of the entropy process corresponds to a subtask accomplishment.
	Our process segmentation algorithm presented in the sequel partitions a response process through identifying the U-shaped curves in the corresponding entropy process.

	Based on the above understanding, we segment a response process in two steps. 
	First, we identify all local maxima of the corresponding entropy process $\bm h$ that are the potential endpoints of U-shaped curves. 
	Then, we filter the set of local maxima and keep those that form a ``deep'' U-shaped curve. 
The response process is then decomposed into several ``deep'' U-shaped curves.
A U-shaped curve is considered to be deep if the entropy change within the curve is significant relative to the entropy fluturation in the entire sequence.
Technically, a subsequence $h_{i:j}$ of $\bm h$ is called a \textit{U-curve} if 
\begin{equation}\label{eq:U_curve_def}
\min \{h_i, h_j\} - \min_{i \leq t \leq j} h_t \geq  \lambda \left( \max_{1 \leq  t \leq T-1} h_t - \min_{1 \leq t \leq T-1} h_t     \right),
\end{equation}
where $\lambda \in [0,1]$ controls the minimum relative depth of a U-shaped curve. 
When $\lambda = 0$, any subprocess between two consecutive local maxima is a U-curve and the sequence will be partitioned into a number of short subsequences. When $\lambda = 1$ and the global maximum of $\bm h$ is unique, no subsequence of $\bm h$ is qualified as a U-curve and the entire sequence is treated as a single subtask. With the above definition, the detailed segmentation procedure is described in Algorithm \ref{al:seg}. 

\begin{algorithm}[Sequence segmentation algorithm]\label{al:seg}
	Given $\lambda$, a response process $\bm s$ is segmented as follows. 
	\begin{enumerate}
		\item Set $h_0 = h_T = \infty$ and $\mathcal{L} = \mathcal{R} = \emptyset$. Initialize $i = 0, j = T$.
		\item Compute entropy process $\boldsymbol{h} = (h_1, \ldots, h_{T-1})$ based on the predictive model of   $\boldsymbol{s} = (s_1, \ldots, s_T)$.
		\item Find  $\mathcal{D}$, the set of local maxima of $\boldsymbol{h}$.
		\item Filter $\mathcal{D}$ according to the following steps to keep the endpoints of $U$-curves. 
		
		~~~~~~~For each $i \in \mathcal D$
		\begin{enumerate}[(i)]
			\item Find the smallest $i' \in \mathcal{D}$ such that $h_{i:i'}$ is U-curve. Add $i'$ to $\mathcal{L}$ and set $i \leftarrow i'$. \label{step2L}
			\item Repeat \eqref{step2L} until $\mathcal{D}$ is exhausted. \label{step2L2}
			
			{For each  $j \in \mathcal D$}
			
			\item Find the largest $j' \in \mathcal{D}$ such that $h_{j':j}$ is U-curve. Add $j'$ to $\mathcal{R}$ and set $j \leftarrow j'$. \label{step2R}
			\item Repeat \eqref{step2R} until $\mathcal{D}$ is exhausted. \label{step2R2}
		\end{enumerate}
		\item Output $\mathcal{L} \cup \mathcal{R}$ as the set of segmentations.
	\end{enumerate}
\end{algorithm}

In Step 4, the local maxima set $\mathcal{D}$ is filtered from left to right in substeps \eqref{step2L}--\eqref{step2L2} and from right to left in substeps \eqref{step2R}--\eqref{step2R2}. When solving a problem, respondents may explore in a wrong direction or take many actions repetitively before figuring out the next subtask.
	In this case, the entropy process remains at a high level for a long period before entering a new U-curve (Subtask EXPLORE in Figure \ref{fig:U_curve}). The bi-directional filtering is used to identify the high-entropy subsequences between two U-curves as an exploratory task.



	Algorithm \ref{al:seg} requires a pre-determined $\lambda$ as input. If $\lambda$ is too small, the processes will be partitioned into small pieces while if $\lambda$ is too large, the processes will be partitioned coarsely. 
	In both situations, the subprocesses are likely to be less interpretable.
	We investigate the effect of $\lambda$ on sequence segmentation in Section \ref{sec:sim}. The results show that segmentation is robust to the choice of $\lambda$ as long as it is away from 0 and 1. We recommend to choose $\lambda$ between 0.2 and 0.8. One can also try different values of $\lambda$ and select the one with the most interpretable results.


\subsection{Subtask clustering}\label{sec:label}

Algorithm 1 segments response processes into subprocesses. It does not, however, reveal how subprocesses are related to subtasks. In this subsection, we develop a clustering method for subprocesses and relate each cluster to a subtask.
Specifically, we convert each subprocess to a fixed-dimensional real-valued vector and then apply standard clustering algorithms such as the k-means clustering \citep{forgy65kmeans,macqueen1967kmeans,arthur07kmeans}.

Recall that Algorithm 1 segments a response process $\bm s$  into $L$ subprocesses at $0 = t_0 < t_1 < \ldots < t_{L} = T$. We associate the $l$-th subprocess $s_{(t_{l-1}+1):t_l}$ with an action frequency profile $\bm z_l = (z_{l1}, \ldots, z_{lM})^\top$. More specifically, the $j$-th element of $\bm z_l$ is the relative frequency of action $a_j$ in the subprocess
\begin{equation}\label{eq:prof_vec}
z_{lj} = \frac{1}{t_l - t_{l-1}} \sum_{t = t_{l-1}+1}^{t_l} \delta_j(s_t), \quad 1 \leq j \leq M, 1 \leq l \leq L,
\end{equation}
where $\delta_j(\cdot)$ is the indicator function defined in \eqref{eq:indicator}. Note that the action frequency profile is a probability vector, namely, $0 \leq z_{lj} \leq 1$, $j = 1, \ldots, M$ and $\sum_{j=1}^M z_{lj} = 1$.
For example, the subprocess ``Start, Toolbar\_Sort, Sort\_A\_2, Sort\_OK" has four actions and each action appears only once. Assuming that these four actions are indexed as $a_1, \ldots, a_4$ in $\mathcal{A}$, then the action frequency profile associated with this subprocess is $ \left(\frac{1}{4}, \frac{1}{4}, \frac{1}{4}, \frac{1}{4},  0, \ldots, 0\right)$.

\begin{figure}
	\centering
	\includegraphics[trim = 0cm 5cm 2cm 0cm, width = \linewidth]{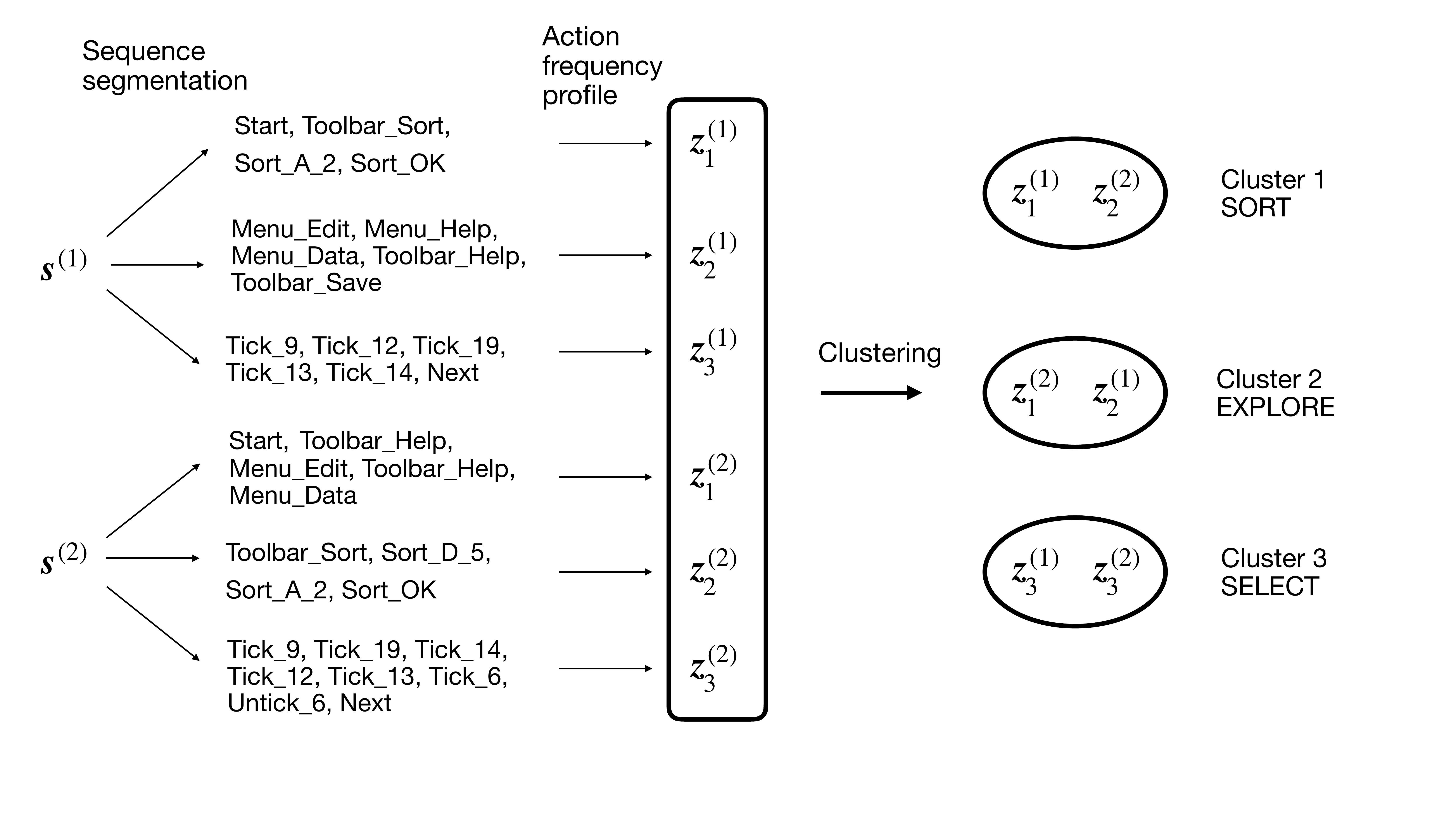}
	\caption{An example of subtask clustering. }\label{fig:clustering}
\end{figure}

For a set of response processes $\mathcal{S} = \{\bm s^{(i)}; i = 1, \ldots, N\}$, let $\bm z_l^{(i)}$ be the action frequency profile associated with the $l$-th subprocess of the $i$-th response process $\bm s^{(i)}$ and $L_i$ be the number of subprocesses for $\bm s^{(i)}$. 
We group similar subprocesses into $R_c$ clusters by performing k-means clustering on $\left\{\bm z_l^{(i)}; i=1, \ldots, N, l = 1, \ldots, L_i\right\}$ with the Hellinger distance. 
The Hellinger distance between two $M$-dimensional probability vectors, $\bm p$ and $\bm q$, is defined by
\begin{equation}
d_H\left(\bm p, \bm q\right) = \sqrt{\sum_{j=1}^M \left(\sqrt{p_j} - \sqrt{q_j}\right)^2}.
\end{equation}
The number of clusters $R_c$ is prespecified and in practice, can be validated by cluster interpretations and prior knowledge of item subtasks.  In our PIAAC data case study, three to seven clusters are often adequate to yield sensible results. After clustering, the interpretation of a cluster can be found by examining the overall action frequency profile of the cluster. Each cluster is then identified as a subtask of the whole problem based on those interpretations. Figure \ref{fig:clustering} illustrates the subtask identification procedure with two response processes $\bm s^{(1)}$ and $\bm s^{(2)}$, where $\bm s^{(1)}$ is the previously mentioned example in the introduction and $\bm s^{(2)}$ is ``Start, Toolbar\_Help, Menu\_Edit, Toolbar\_Help, Menu\_Data, Toolbar\_Sort, Sort\_D\_5, Sort\_A\_2, Sort\_OK, Tick\_9, Tick\_19, Tick\_14, Tick\_12, Tick\_13, Tick\_6, Untick\_6, Next ". The second response process differs from the first one  in the order of subtasks as well as actions used within a subtask.

	Let $\mathcal{G} = \{g_1, \ldots, g_R\}$ be the set of identified subtasks/clusters. For the item described in the introduction, $\mathcal{G} = \{\text{EXPLORE}, \text{SELECT}, \text{SORT}\}$. Our task identification procedure associates each action $s_t$ in a response process $\bm s$ with a subtask $ v_t  \in \mathcal{G}$. 
	We therefore obtain a new process $ \hat{\bm v} = \left\{\hat v_1, \ldots, \hat v_T\right\}$. 
 Since actions in the same subprocess of $\bm s$ come from the same subtask, we can remove all consecutively repeated $\hat v_t$ from $\hat{\bm v}$ to obtain another sequence $\hat{\bm q}$. 	We call $\hat{\bm v}$  the state sequence and $\hat { \bm q}$  the subtask sequence. In the example illustrated by Figure \ref{fig:clustering}, the subtask sequence of $\bm s^{(1)}$ is ``SORT, EXPLORE, SELECT" and the subtask sequence of $\bm s^{(2)}$ is ``EXPLORE, SORT, SELECT".
	The subtask sequence is often much shorter than the original response process and have less but more informative variations. 
	These subtask sequences are helpful for visualizing and analyzing respondents' problem-solving strategies. We will illustrate it through a real data analysis.


We conclude the section with a summary of the proposed subtask identification procedure (SIP) in Procedure \ref{proc:SIP} below.


\begin{procedure}[Subtask identification procedure]\label{proc:SIP}
	SIP consists of the following three steps.
	\begin{enumerate}
		\item (Prediction step) Fit an action predictive model and obtain the predictive distributions.
		\item (Segmentation step) Partition the response process into multiple subprocesses based on the corresponding entropy process according to Algorithm \ref{al:seg}.
		\item (Labeling step) Cluster the subprocesses according to their action frequency defined in \eqref{eq:prof_vec} and label the clusters as subtasks.
	\end{enumerate}
\end{procedure}

\section{Simulations}\label{sec:sim}
\subsection{Data generation}

We simulate response processes of a problem-solving item with four subtasks ($R = 4$) and 26 possible actions ($M = 26$). 
The problem-solving subtasks are denoted by upper case letters $A$, $B$, $C$, $D$, i.e. subtask set $\mathcal{G} = \left\{A, B, C, D\right\}$; the actions are denoted by lower case letters, i.e. action set $\mathcal{A} = \left\{a, b, \ldots, z\right\}$. 
To obtain a simulated response process, we first generate a subtask sequence $\bm q = \left(q_1, \ldots, q_L\right)$, where $q_l \in \mathcal{G}$, and then generate, for each $l$, an action sequence $\tilde{\bm s}_l$ as a subprocess in subtask $q_l$. The full simulated response process is then obtained by concatenating the $L$ action sequences, namely, $\bm s = \left(\tilde{\bm s}_1, \ldots, \tilde{\bm s}_L\right)$. Both the subtask sequence $\bm q$ and the subprocesses $\tilde{\bm s}_l$, $l = 1, \ldots, L$, are generated from Markov models; see Appendix \ref{sec:appendix:sim} for details.




\subsection{Experiment settings}\label{sec:sim_settings}
We generate 100 datasets, each of which contains $N=5000$ response processes according to the data generation procedure. 
For each dataset, $N$ action sequences are randomly partitioned into training (70\%), validation (15\%) and test (15\%) sets. We train the RNN model for 50 epochs with validation-based early stopping. The latent dimension $K$ is set to be 20 and the learning rate of RmsProp optimizer is $10^{-3}$. 
Nine values of $\lambda$, $0.1, 0.2, \ldots, 0.9$, are used in the segmentation step to examine the sensitivity of subtask segmentation results to $\lambda$. For each $\lambda$ value, the segmentation algorithm is applied to all the processes in the dataset. In the labeling step, the number of clusters is set to the true number of subtasks $R = 4$. The samples in the training and validation sets are used to perform the k-means clustering. 
The segmented subprocesses in the test set are then assigned to the cluster whose centroid is the closest to the action frequency profile of the subprocess in terms of the Hellinger distance. 

As a comparison, we also fit a hidden Markov model \citep[HMM;][]{baum1966hmm}. The estimated hidden states produced by HMM are treated as the estimated subtask sequence in this case. The number of hidden states in HMM is also set to be the true number of subtasks.

\subsection{Evaluation Criteria}\label{sec:eval}



The evaluation criteria described below are used for both the simulated and real data analysis. For notational simplicity, the superscript  of the response process is omitted when there is no ambiguity.
Let $\bm v$ and $\hat{\bm v}$ be the true and the estimated state sequences, respectively.
	We use $\mathcal{T} = \left\{t: v_{t+1} \neq v_t,  t < T\right\}$ to denote the set of times immediately before a transition. 
	Similarly, we write $\hat{\mathcal{T}} = \left\{t: \hat v_{t+1} \neq \hat v_t,  t < T\right\}$.
Hence, $\mathcal{T} \cap \hat{\mathcal{T}}$ represents the set of correctly estimated subtask transitions. We define Precision and Recall for subtask in Table \ref{table:measures}. Note that they are different from the precision and recall for action prediction.
To further take into account the direction of the transitions, we use
 $\mathcal{T}_{+} =  \left\{t : v_{t+1} \neq v_t, \hat v_{t+1} \neq \hat v_t, \hat v_{t+1} = v_{t+1},  t < T\right\}$ to denote the set of times where both the transition time and the label of the next subtask are correct. The precision and recall corresponding to $\mathcal{T}_{+}$ are denoted by Precision+ and Recall+, respectively. The perfect match (overlap) between $\bm v$ and $\hat{\bm v}$ is denoted by Overlap, and we write $\mathcal{V} = \left\{t : v_t = \hat v_t \right\}$. The exact mathematical definitions (formulas) are given in Table \ref{table:measures}. 


\begin{table}[htb]
	\centering
	\caption{Five measures for comparing the estimated and the true state sequences.}\label{table:measures}
	\begin{tabular}{cc}
		\hline
		Measures & Formulas \\
		\hline
		Precision & $\sum_{i \in \Omega} | \mathcal{T}^{(i)} \cap \hat{\mathcal{T}}^{(i)} | / \sum_{i \in \Omega}| \hat{\mathcal{T}}^{(i)} | $\\
		Recall & $\sum_{i \in \Omega} | \mathcal{T}^{(i)} \cap \hat{\mathcal{T}}^{(i)} | / \sum_{i \in \Omega}| \mathcal{T}^{(i)} | $ \\
		Precision+ & $\sum_{i \in \Omega} | \mathcal{T}^{(i)}_{+}  | / \sum_{i \in \Omega}| \hat{\mathcal{T}}^{(i)} | $\\
		Recall+ & $\sum_{i \in \Omega} | \mathcal{T}^{(i)}_{+}  | / \sum_{i \in \Omega}| \mathcal{T}^{(i)} | $\\
		Overlap & $\sum_{i \in \Omega} |\mathcal{V}^{(i)}| / \sum_{i \in \Omega} T_i $ \\
		\hline
	\end{tabular}
\end{table}

\subsection{Results}
\begin{figure}[ht]
	\centering
	\includegraphics[width=\textwidth]{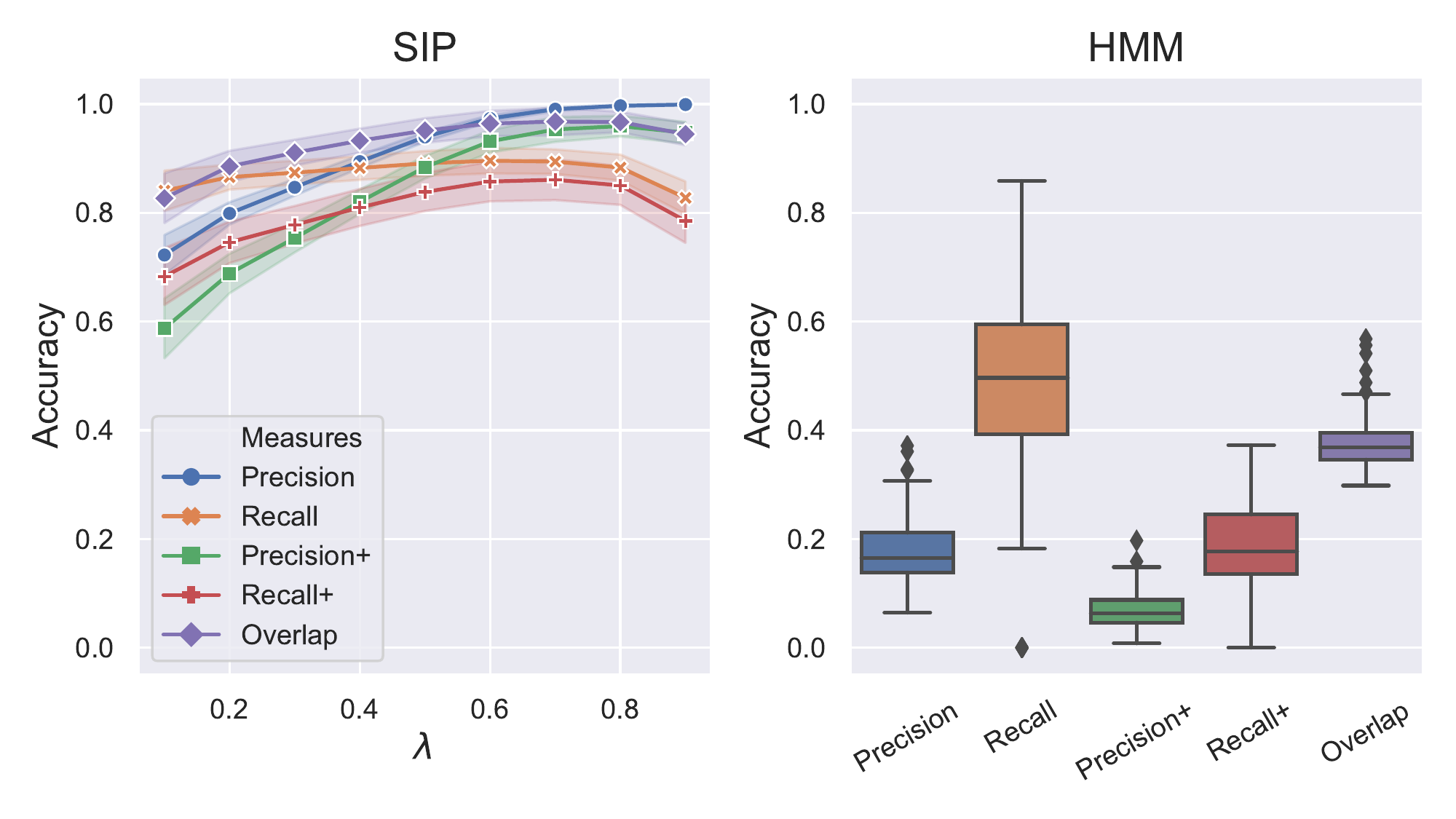}
	\caption{Accuracy on subtask identification. Left: Mean accuracy of SIP under five measures and threshold values from 0.1 to 0.9. The standard errors are shown by error bands. Right: Accuracy of HMM under five measures.}
	\label{fig:sim_transition_acc}
\end{figure}

Figure \ref{fig:sim_transition_acc} summarizes the results for subtask identification. The five measures of subtask estimation accuracy defined in Table \ref{table:measures} are plotted against the values of $\lambda$ in the left panel of the figure. 
When $\lambda$ is small, Precision and Precision+ are relatively low since the processes are overly segmented and unnecessary partitions are produced; when $\lambda$ is too large (greater than 0.8), Recall and Recall+ decrease since the action sequences are under segmented and many subtasks are not captured. 
The performance is robust to the choice of $\lambda$ when its value stays away from 0 and 1. 
The right panel of Figure \ref{fig:sim_transition_acc} displays boxplots of the five measures for the estimated state sequences obtained from HMM on the 100 datasets. 
The Precision, Precision+, Recall+, and Overlap are all below 0.6. The Recall has a median around 0.5 and a wide spread. 
It is clear that the proposed SIP performs substantially better, noting that the vertical axes of the two panels are on the same scale.


\section{Case Study}\label{sec:case_study}
In this section, we present a case study of the process data from PIAAC 2012.
We investigate the performance of SIP on PIAAC 2012 and demonstrate how the results can be exploited to study respondents' problem-solving strategies and their relationship with other variables. In Section \ref{sec:example_U23}, we examine the accuracy of subtask identification.
We show in Section \ref{sec:visual} that the segmentation helps visualize respondents' problem-solving strategies. The relationship between problem-solving strategy and efficiency is analyzed in Section \ref{sec:strategy}. 
The threshold $\lambda$ in SIP is set to 0.3 for all the items.

\subsection{Data description}
The dataset contains log files involving 14 PSTRE items in PIAAC 2012. There are $40,230$ respondents from 17 countries. Each respondent answered all or a subset of the 14 items. The required tasks in the items are diverse, ranging from email classification, spread sheet handling, web browsing to scheduling, etc. The number of possible actions varies from 27 to 1536 across items. 
Those response processes with fewer than ten actions are excluded as short processes are usually the result of inattentive responses that provide little information of respondents' problem-solving behaviors.
The dataset also includes the final response outcomes for more than 97\% of the response processes. For some items, the original response outcomes are polytomous. We simplify them to dichotomous outcomes by labeling the fully correct responses as 1 and 0 otherwise. Some basic statistics of the preprocessed data are listed in Table \ref{table:data} for each item, where $N$ is the number of respondents, $M$ denotes the number of possible actions and $\bar T$ stands for the average process length. ``Correct \%'' is the percentage of correct responses among all recorded response outcomes. 
Although our method can be applied to all items, only three items, U23, U01b and U19a,  are selected to demonstrate the performance of the proposed subtask identification procedure and how the subtask sequences can help understand respondents' problem-solving behaviors.

\begin{table}[htb]
	\centering
	\begin{threeparttable}
		\caption{Descriptive statistics of 14 PIAAC problem-solving items.}\label{table:items}
		\begin{tabular}{llcccc}
			\hline
			ID & Description & $N$ & $M$ & {\small $\bar T$} & Correct \% \\
			\hline
			U01a & Party Invitations - Can/Cannot Come & 20930 & 66 & 21.2 & 72.8\\
			U01b & Party Invitations - Accommodations & 20859 & 60 & 32.0 & 63.6\\
			U02 & Meeting Rooms & 17404 & 102 & 36.7 & 19.8\\
			U03a & CD Tally & 8798 & 67 & 15.5 & 84.1\\
			U04a & Class Attendance & 16498 & 1536 & 60.0 & 21.7\\
			U06a & Sprained Ankle - Site Evaluation Table & 8034 & 30 & 13.9 & 28.6\\
			U06b & Sprained Ankle - Reliable/Trustworthy Site & 17307 & 27 & 19.8 & 57.2\\
			U07 & Digital Photography Book Purchase & 16172 & 41 & 24.2 & 71.2\\
			U11b & Locate E-mail - File 3 E-mails & 17085 & 140 & 33.5 & 31.9\\
			U16 & Reply All & 20704 & 896 & 39.4 & 74.1\\
			U19a & Club Membership - Member ID & 19796 & 197 & 21.2 & 84.8\\
			U19b & Club Membership - Eligibility for Club President & 16821  & 562 & 28.8 & 68.1\\ 
			U21 & Tickets & 6430 & 149 & 22.2 & 45.1\\
			U23 & Lamp Return & 18730 & 177 & 27.6 & 50.6\\
			\hline
		\end{tabular}\label{table:data}
		\begin{tablenotes}
			\item Note: $N$ = number of respondents; $M$ = number of possible actions; $\bar T =$ average process length; Correct \% = percentage of correct responses. Data has been cleaned by merging consecutive and removing redundant actions. Processes with fewer than 10 actions are excluded.
		\end{tablenotes}
	\end{threeparttable}
\end{table}

\subsection{Identification accuracy} \label{sec:example_U23}

For each item discussed in the case study, we randomly partition the response processes into training (70\%), validation (15\%), and test (15\%) sets, denoted by $\mathcal{S}_{\text{train}}$, $\mathcal{S}_{\text{valid}}$, and $\mathcal{S}_{\text{test}}$, respectively. 
The RNN-based action prediction model is trained for $50$ epochs on $\mathcal{S}_{\text{train}}$. To avoid overfitting, we monitor the log likelihood $l\left(\bm \eta; \mathcal{S}_{\text{valid}}\right)$ and take a validation-based early stopping. The latent dimension $K$ is 20 and the learning rate is $10^{-3}$.

We use item U23 as an example to demonstrate the accuracy of subtask identification. 
For this item, respondents were asked to request an exchange for a desk lamp purchased online because of a wrong color. 
To complete the task, respondents need to finished several subtasks involving email and web browsing.
In the subtask clustering step, the number of clusters is set to seven ($R_c = 7$). After clustering, the interpretation of each cluster may be ascertained by examining the high frequency actions in the cluster. Two of the clusters are related to clicking different links on the web page that are irrelevant to the required task. Therefore, they are merged into a single subtask EXPLORE\_LINK. Each of the remaining clusters is identified as a single subtask. The six resulting subtasks and their interpretations are listed in Table \ref{U23}.

\begin{table}[ht]
	\centering
	\caption{Problem-solving subtasks in item U23.}\label{U23}
	{\small
		\begin{tabular}{c c p{.50\textwidth}}
			\hline
			Subtask & Proportion* & Interpretation\\
			\hline
			EXPLORE\_MAIL & 27.3\% &  View and handling emails. \\
			EXPLORE\_LINK & 22.9\% & Exploration of links and boxes.  \\
			SUBMIT& 16.9\% &  Submit form and proceed to the next item. \\
			OBTAIN\_CODE & 14.0\% & Click button to obtain authorization code. \\
			ENTER\_CODE & 9.6\% & Enter correct authorization code. \\
			FILL\_FORM & 9.3\% & Fill the reason and/or objective in return form. \\
			\hline
		\end{tabular}
    }
	\begin{tablenotes}
		\item *Proportion = the aggregated number of actions within each subtask, divided by the total number of actions. 
	\end{tablenotes}
\end{table}

Evaluating the accuracy of subtasks  identified by SIP requires the ground truth of the state sequences $\bm v$ for at least a subset of the response processes. To obtain such reference sequences, we randomly sample 100 response from $\mathcal{S}_{\text{test}}$, denoted by $\widetilde{\mathcal{S}}_{\text{test}}$, and manually label them according to the six identified subtasks. These resulting state sequences and the corresponding subtask sequence are treated as the truth and we compare the estimated ones with them for evaluation.

\begin{figure}[htb]
	\centering
	\includegraphics[width=\textwidth]{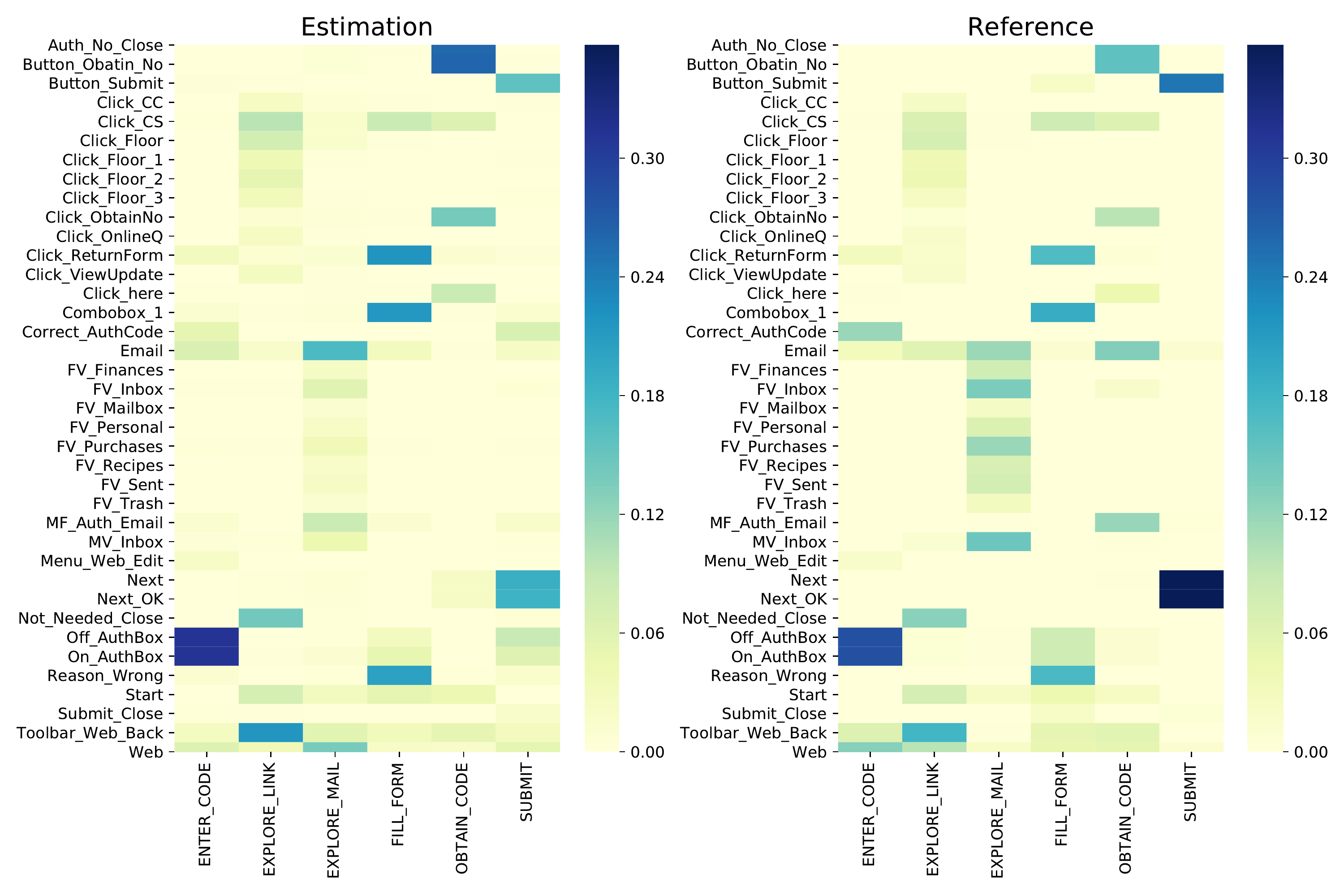}
	\caption{Action frequency profiles for the estimated and true state sequences of U23. Actions that appear fewer than 2\% of the total action occurrences are omitted in the plot.
	}
	\label{fig:action_freq_subtask}
\end{figure}

To verify the interpretation of the identified subtasks from our procedure, we compare the action frequency profile for the estimated and true state sequences. These profiles are exhibited in Figure \ref{fig:action_freq_subtask}, where the left and right panels correspond to the estimated and true action frequency profiles, respectively. Each column represents a probability vector that sums to one. 
Two observations can be made from the figure.
First, the  identified subtasks can be well-distinguished by their action frequency profiles. The high frequency actions under each subtask are closely related to the subtask accomplishment.
Second, the action frequency profiles in the left and right panels are similar, suggesting that the proposed procedure can reasonably identify the problem-solving subtasks.

\begin{figure}[htb]
	\centering
	\includegraphics[width=\textwidth]{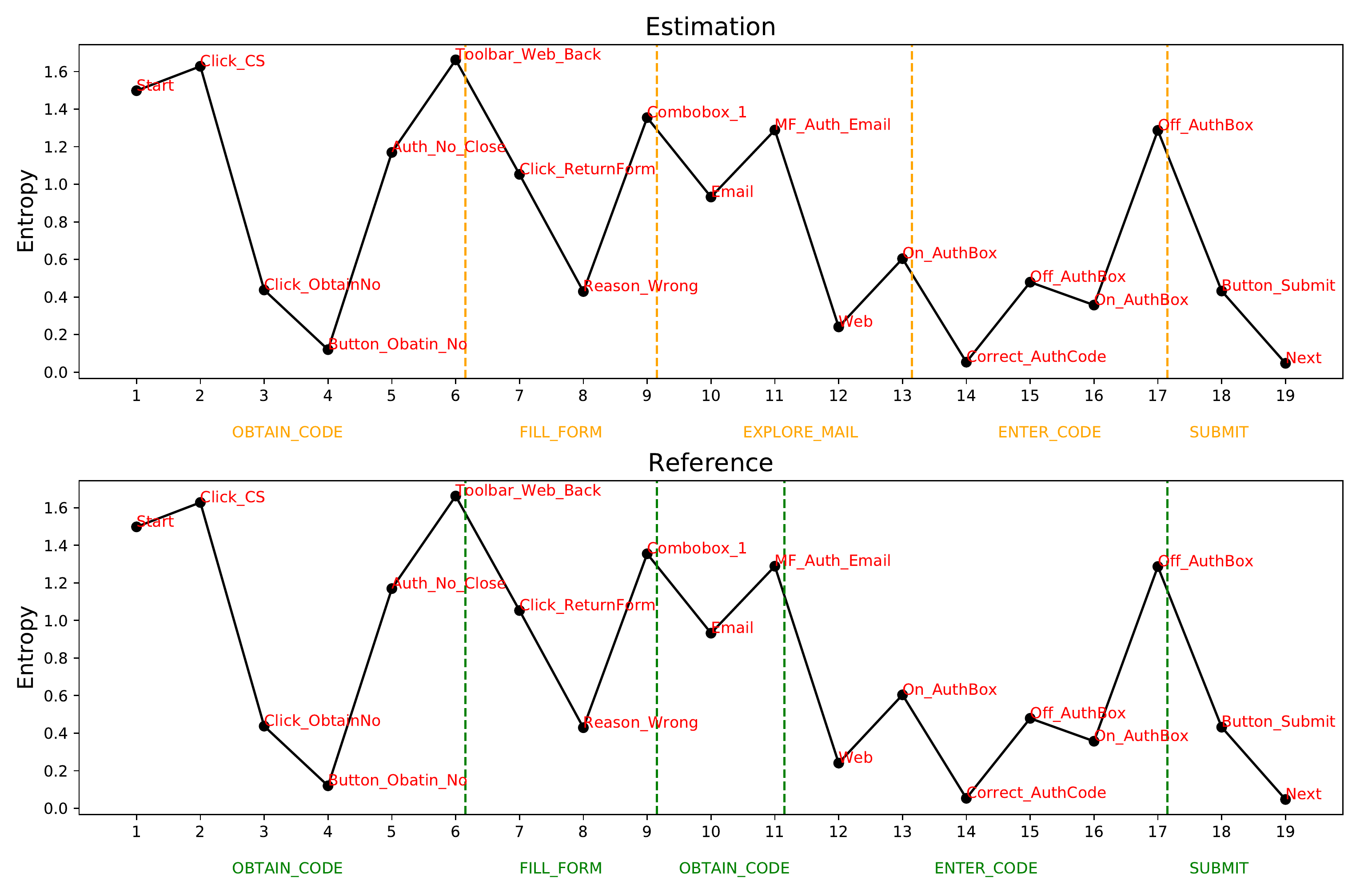}
	\caption{Subtask identification result of a typical response process from U23. Top: Problem solving subtasks identified by SIP. Bottom: Manually labeled problem-solving subtasks.}
	\label{fig:subtask_seg_ex}
\end{figure}

Figure \ref{fig:subtask_seg_ex} shows the subtask identification result for a typical process in $\widetilde{\mathcal{S}}_{\text{test}}$. 
The solid lines are the entropy process derived from the fitted RNN model in the one-step-ahead prediction. In the top panel, the dashed lines stand for the segmentation locations identified by the SIP. The identified subtask of each subprocess is marked below the horizontal axis. In the bottom panel, the dashed lines and the labeled subtasks are obtained from the true state sequence. 
For this process, SIP correctly identified most of the subtasks in terms of both the transition location and category. The only exception is that the third identified subtask has a longer subprocess than the reference and is incorrectly labeled as EXPLORE\_MAIL.

We evaluate the accuracy of subtask identification on $\widetilde{\mathcal{S}}_{\text{test}}$ by the five criteria defined in Table \ref{table:measures} and the results are summarized in Table \ref{table:eval_SIP_HMM}. 
We also include the results from a HMM in the table for comparison. The estimated hidden states are treated as the estimated state sequence for HMM.
It is clear from Table \ref{table:eval_SIP_HMM} that SIP produces more balanced Precision and Recall, which measure the accuracy of estimated subtask locations. It also gives more balanced Precision+ and Recall+, which take into account both the transition location and direction. In contrast, HMM has higher Recall and Recall+, but lower Precision and Precision+. It tends to divide a process into shorter and less interpretable subprocesses. The estimated state sequence from SIP also has a higher Overlap than that from HMM.
\begin{table}[htb!]
	\centering
	\caption{Evaluation of subtask results.}\label{table:eval_SIP_HMM}
	\begin{tabular}{cccccc}\hline
		& Precision & Recall & Precision+ & Recall+ & Overlap \\
		\hline 
		SIP & 0.335 & 0.327 & 0.251 & 0.246 & 0.659 \\
		HMM & 0.200 & 0.859 & 0.072 & 0.309 & 0.330 \\ \hline
	\end{tabular}
\end{table}



\subsection{Subtask visualization}\label{sec:visual}
The detailed action information in response processes makes process data an important source for studying problem-solving behaviors. 
At the same time, the excessive length of the sequences and the high variability in the sequence elements make it difficult to visualize the detailed processes to intuitively understand the problem-solving process.
With SIP, a response process is simplified into a much shorter and less variable subtask sequence while keeping the primary problem-solving steps.
By visualizing these subtask sequences, the similarities and differences among responses can be detected easily.


We use item U01b to demonstrate how to visualize subtask sequences. In U01b, respondents were asked to organize email responses to a party invitation in an email client. New folders should be created to keep track of the attendants' accommodation needs. We identified three problem-solving subtasks (listed in Table \ref{table:U01b}) from the response processes. 

\begin{table}[ht]
	\centering
	\caption{Problem-solving subtasks in item U01b}\label{table:U01b}
	{\small
		\begin{tabular}{c c l}
			\hline
			Subtask & Proportion & Interpretation\\
			\hline
			CREATE\_NEW\_FOLDER & 23.5\% &  Create a new folder to classify emails.\\
			VIEW\_TOOLBAR\_MOVE & 28.5\% &  View email, use toolbar to move it.\\
			VIEW\_DRAG\_DROP & 48\% &  View email, drag and drop the icon to a folder.\\
			\hline
		\end{tabular}
	}
\end{table}

\begin{figure}[htb!]
	\centering
	\includegraphics[width=\textwidth]{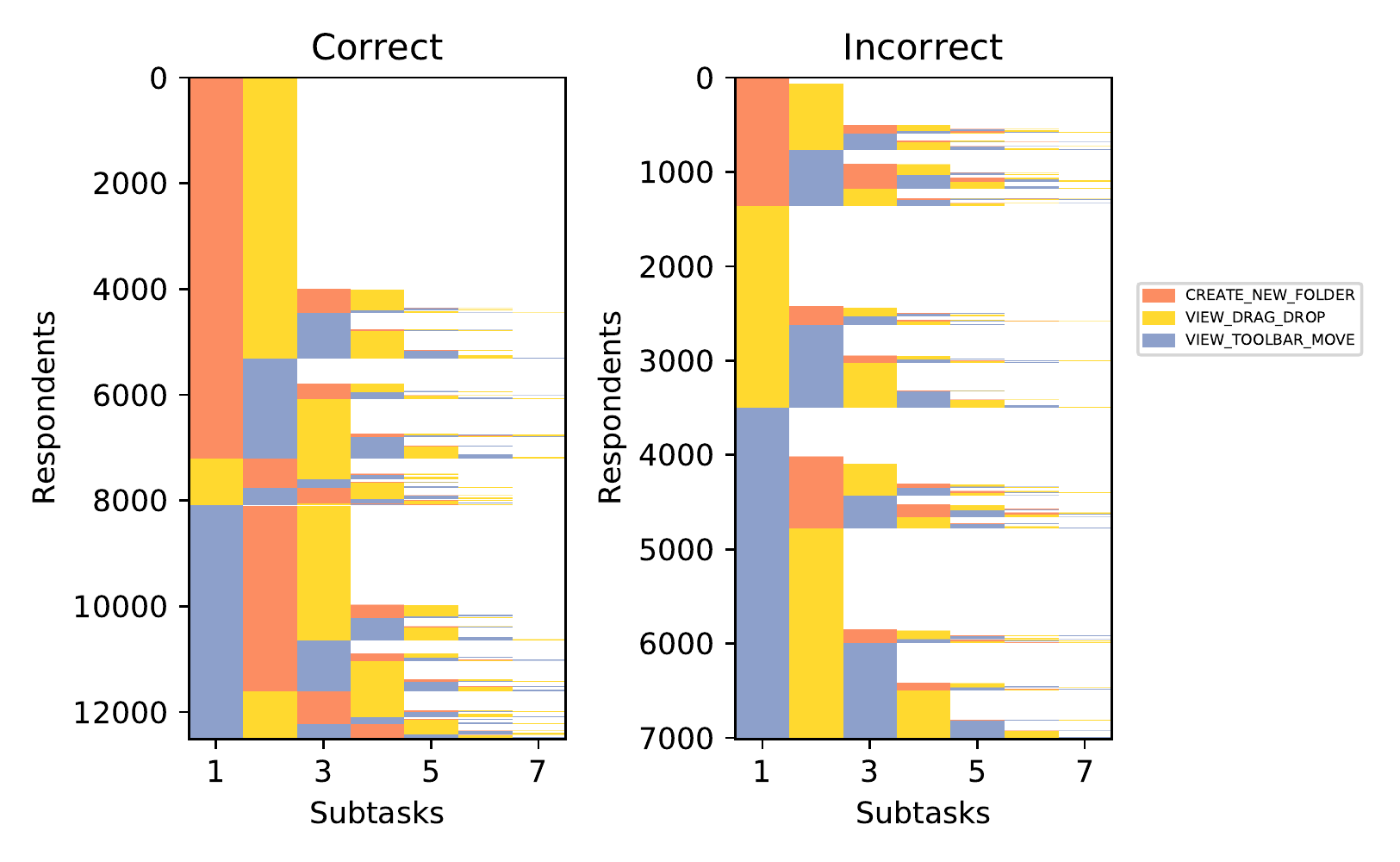}
	\caption{Visualization of the subtask sequences of item U01b. }
	\label{fig:visualize}
\end{figure}

We divide the respondents of U01b into two groups according to their final response outcomes and visualize the subtask sequences for each group in Figure \ref{fig:visualize} as follows.
The three subtasks are represented by three distinct colors in the figure. A subtask sequence $\hat{\bm q}$ of length $L$ is displayed as a horizontal line segment with $L$ units in length. Each unit corresponds to a subtask in the sequence and is colored accordingly. The line segments are then sorted and stacked vertically in the dictionary order of the subtask sequences.
Figure \ref{fig:visualize} presents all subtask sequences with length no greater than seven, representing more than 95\% of the respondents of U01b. 
There are about 12,000 respondents in the correct group and about 7,000 respondents in the incorrect group. By visualizing their subtask sequences, the difference in the problem-solving processes between the groups becomes obvious: most of respondents in the incorrect group did not create new folders, which is an essential step for successfully completing the task.

\subsection{Problem-solving strategies}\label{sec:strategy}
In this section, we use the subtask sequences of item U19a obtained from the SIP to explore the relationship between problem-solving strategies and the final outcome as well as the response time. 
In item U19a, respondents are asked to find out the ID number of a bike club member from a given spreadsheet and then email it to the secretary. 
Using SIP, we identify five problem-solving subtasks as listed in Table \ref{table:U19a}.
We find two strategies after visualizing the subtask sequences. In the first one, the ID number is obtained by searching the name of the club member in the spreadsheet. In the second one, respondents first sort the members in spreadsheet according to their names and then look up the given name to obtain the ID number. 
If the subtask sequence contains SEARCH but not SORT, then it indicates one strategy, whereas if it contains SORT but not SEARCH, then it indicates the other strategy.
In addition, subtask sequences of some respondents contain both SEARCH and SORT, indicating a mixing of the two strategies.
\begin{table}[htb]
	\centering
	\caption{Problem-solving subtasks in item U19a.}\label{table:U19a}
	{\small
		\begin{tabular}{c c l}
			\hline
			Subtask & Proportion & Interpretation\\
			\hline
			EXPLORE & 18.9\% &  Make exploratory actions on email and spreadsheet environments.\\
			SEARCH& 13.4\% &  Use searching tools to find ID number.\\
			SORT& 22.9\% &  Use sorting tools to find ID number. \\
			WRITE\_EMAIL & 17.1\% & Write an email that includes the number founded. \\
			SEND\_EMAIL& 27.7\% & Send email to secretary and continue to the next item. \\
			\hline
		\end{tabular}
	}
\end{table}

The overall percentage of correct responses in U19a is $84.8\%$. We find that respondents can get the correct answer more easily with the SEARCH strategy than the SORT strategy. There are 3,865 respondents who used searching tools only and $91.6\%$ of them solved the problem correctly. Among the 9,812 respondents who used sorting tools only, the proportion of correct responses is $83.1\%$. 

To compare the problem-solving efficiency of the two strategies, we plot the histograms of the logarithm of the response time for the two groups in the top panel of Figure \ref{fig:search_sort_densities}. It shows that respondents in the SEARCH group tend to use less time to finish the item than those in the SORT group.
To further look into the age effect, we display the joint kernel density estimates of respondents' age and response time for the two groups in the lower two plots of Figure \ref{fig:search_sort_densities}. The marginal densities for age and response time are also shown by curves on the top and right side of each plot. 
For a given age, the log response time in the SORT group tends to be longer than that in the SEARCH group.
We also observe that the marginal distribution of age for the SORT group has a heavier tail than that for the SEARCH group, indicating that elder respondents are more inclined to perform SORT than the younger ones. 
In addition, the correlation between age and log response time is $0.277$ for the SEARCH group and $0.329$ for the SORT group. The positive correlations suggest that the elder generally spent more time when applying the same strategy.
\begin{figure}[htb!]
	\centering
	\includegraphics[width=0.44\linewidth]{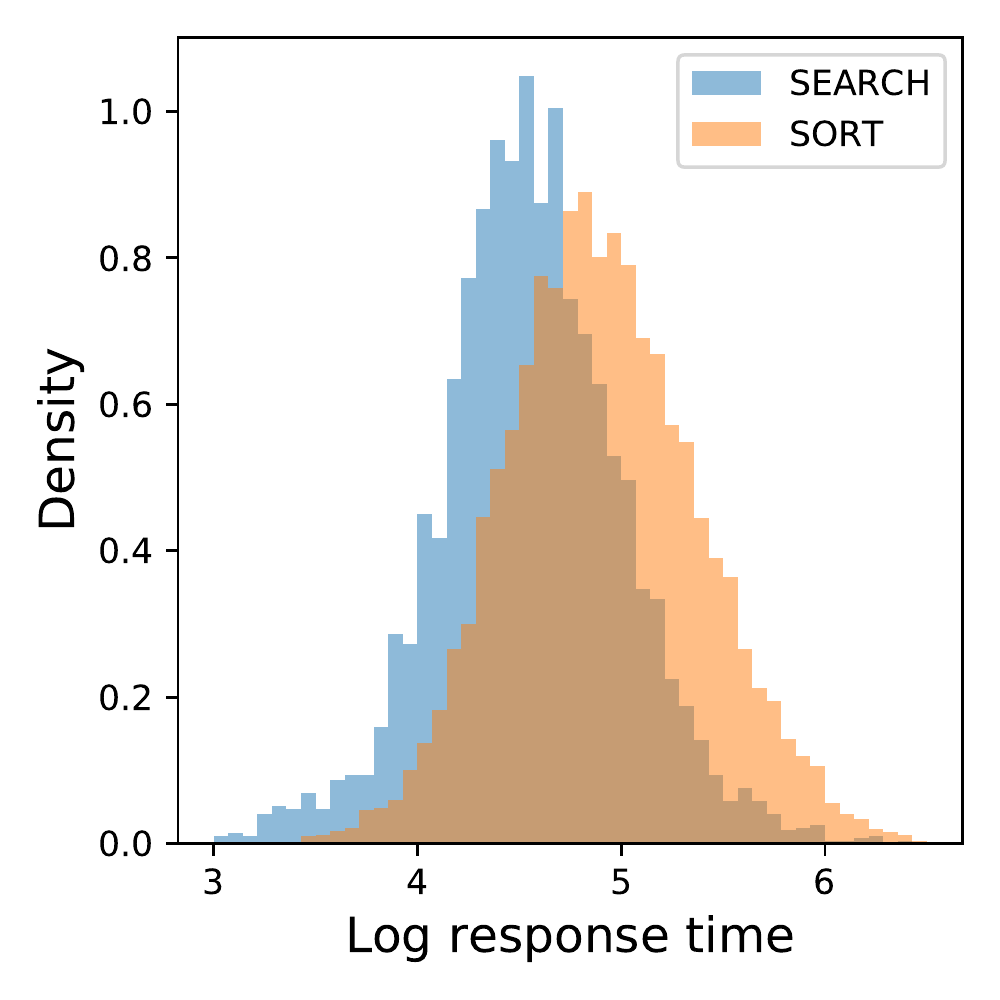}
	\caption{Response time}
\end{figure}
\begin{figure}[htb!]
	\centering
	\begin{minipage}[b]{.49\textwidth}
		\centering
		\includegraphics[width=1\linewidth]{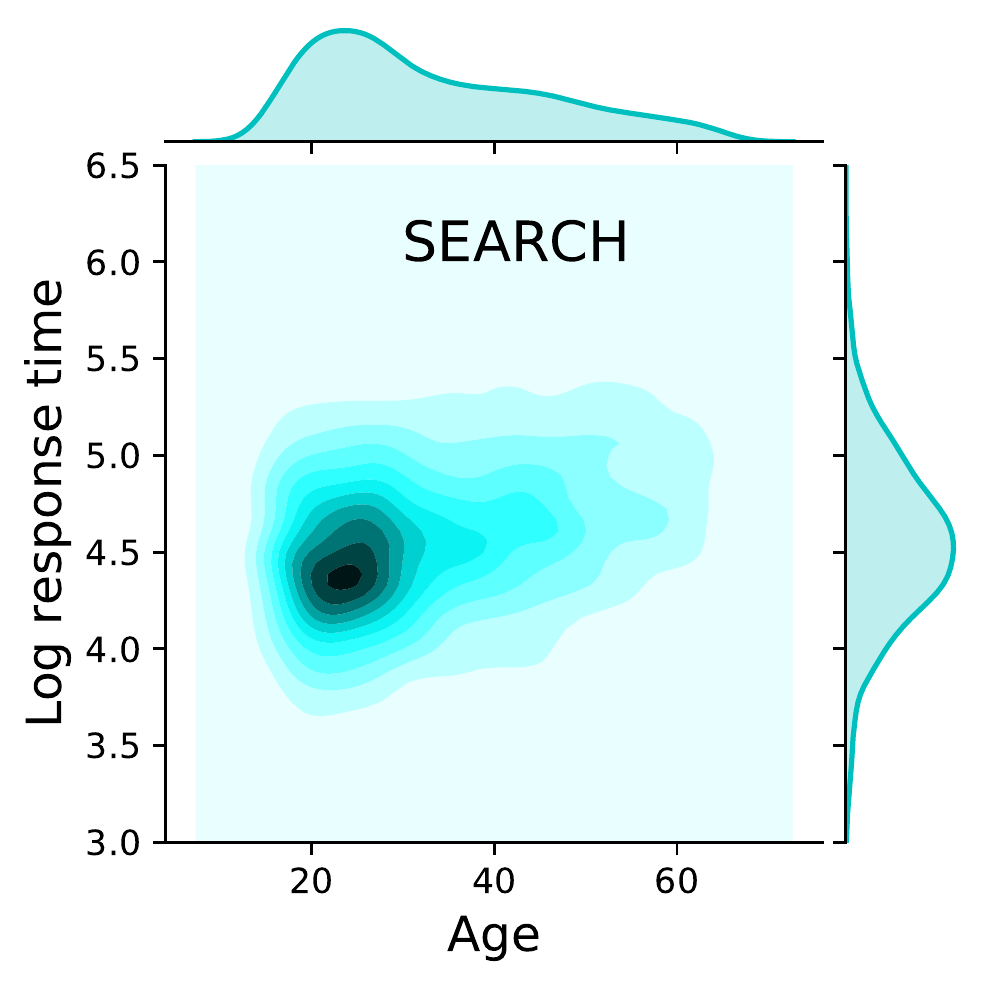}
	\end{minipage}
	\begin{minipage}[b]{.49\textwidth}
		\centering
		\includegraphics[width=1\linewidth]{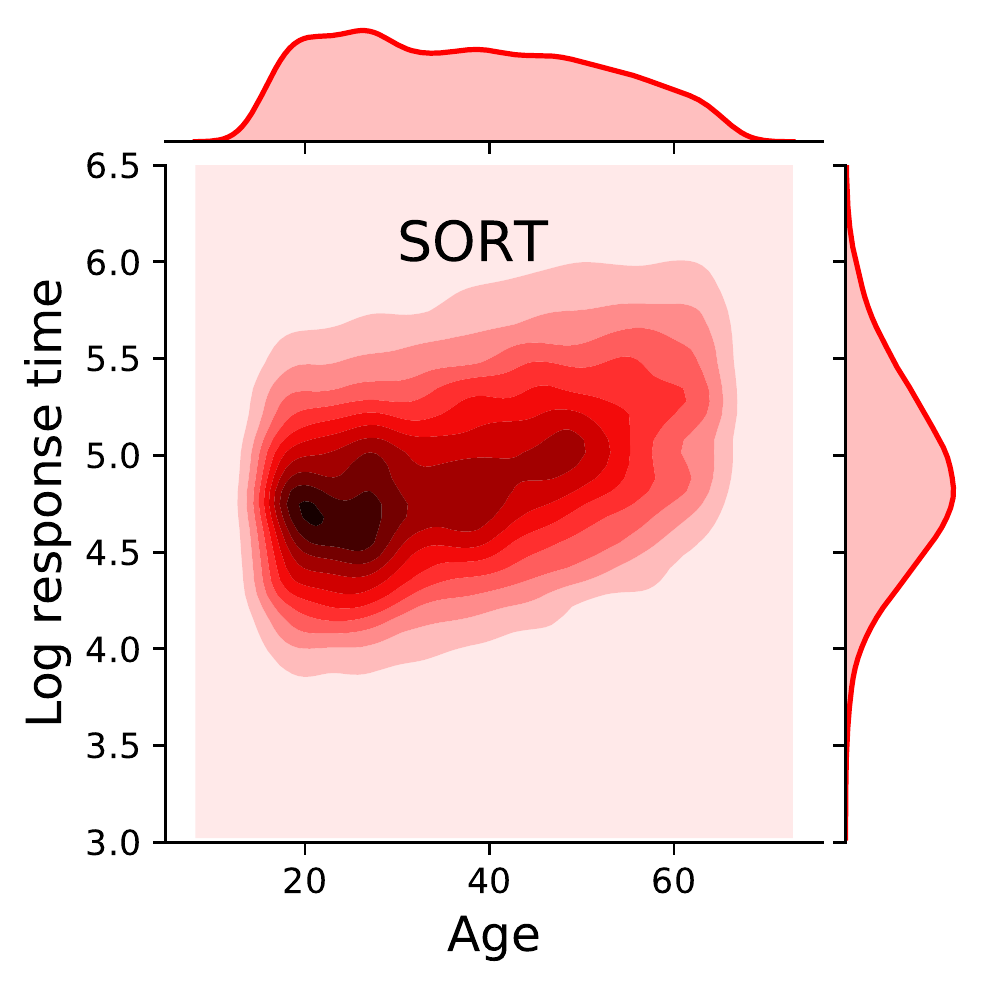}
	\end{minipage}
	\caption{Density plots on response time of item U19a under two strategies.}
	\label{fig:search_sort_densities}
\end{figure}

\subsection{Process information decomposition}
Subtask sequences reflect respondents' high-level problem-solving strategies and thus carry more information than the traditional item responses. However, they also  lose some information compared to the original response processes since detailed actions within each subtask are ignored. 
Information in a response process can be decomposed into two parts, information in the subtask sequence and information within each problem-solving subtask. 

To illustrate the decomposition, we use item U19a as an example and conduct an experiment to predict the final outcome, literacy score, numeracy score and the age of respondent from the information in the response processes. For each target variable, a generalized linear model 
$$g(E(Y)) = \beta_0 + \bm \beta^\top \bm x$$ is considered, where $Y$ denotes the target resposne variable, $g$ is the link function, and $\bm x$ is a feature vector of the response processes of U19a. 
We use a logit link $g(p) = \log(\frac{p}{1-p})$ for the model of the binary outcome and the identity link $g(x) = x$ for the models of other target variables.
To compare the information in the subtask sequence and within subtasks, four different choices of the feature vector $\bm x$ are considered.
\begin{enumerate}[(i)]
\item The feature vector $\bm x$ consists of only the binary outcomes of U19a. We call it the baseline model.
\item Subtask transition features are included in $\bm x$ in addition to the binary outcome. A subtask feature is a binary indicator of whether a transition occurs in the estimated subtask sequence $\hat{\bm q}$. 
\item Unigram and bigram features for each subtask in $\mathcal{G}$ are included in $\bm x$ in addition to the features used in choice (ii). In particular, we only consider those unigrams and bigrams whose frequencies are higher than 0.1\%. If subtask $i$ exists in $\hat{ \bm q}$, then the feature corresponding to a unigram/bigram is 1 if the unigram/bigram appears in the subprocesses with state $i$ and 0 otherwise. If subtask $i$ does not appear in $\hat{ \bm q}$, we set all unigram and bigram features corresponding to the subtask to zero. 
\item Unigram and bigram features for the response process are used in $\bm x$ in addition to the binary outcomes. The feature corresponding to a unigram/bigram is 1 if the unigram/bigram appears in the response process and 0 otherwise. Same as in choice (iii), only unigrams and bigrams whose frequencies are higher than 0.1\% are considered. 
\end{enumerate}
When the target variable is the binary outcome, we do not consider choice (i) and remove the binary outcome from the features in choices (ii)-(iv). To investigate the information contained in a specific subtask $i$, we only consider respondents whose subtask sequences contain $i$ and extract unigram/bigram features from the corresponding subprocesses in the same way as choice (iii).

The model is fitted on $\mathcal{S}_{\text{train}} \cup \mathcal{S}_{\text{valid}}$ for each choice of feature vector and the prediction performance is evaluated on $\mathcal{S}_{\text{test}}$. The evaluation criterion is the area under the ROC curve (AUC) for binary outcome and the out-of-sample $R^2$ ($\text{OSR}^2$) for other variables. To avoid overfitting, $L_2$ penalties are placed on the coefficients for both the logistic and the linear model. The penalty parameter is determined by five-fold cross-validation. 

\begin{figure}[htb!]
	\centering
	\includegraphics[scale=0.7]{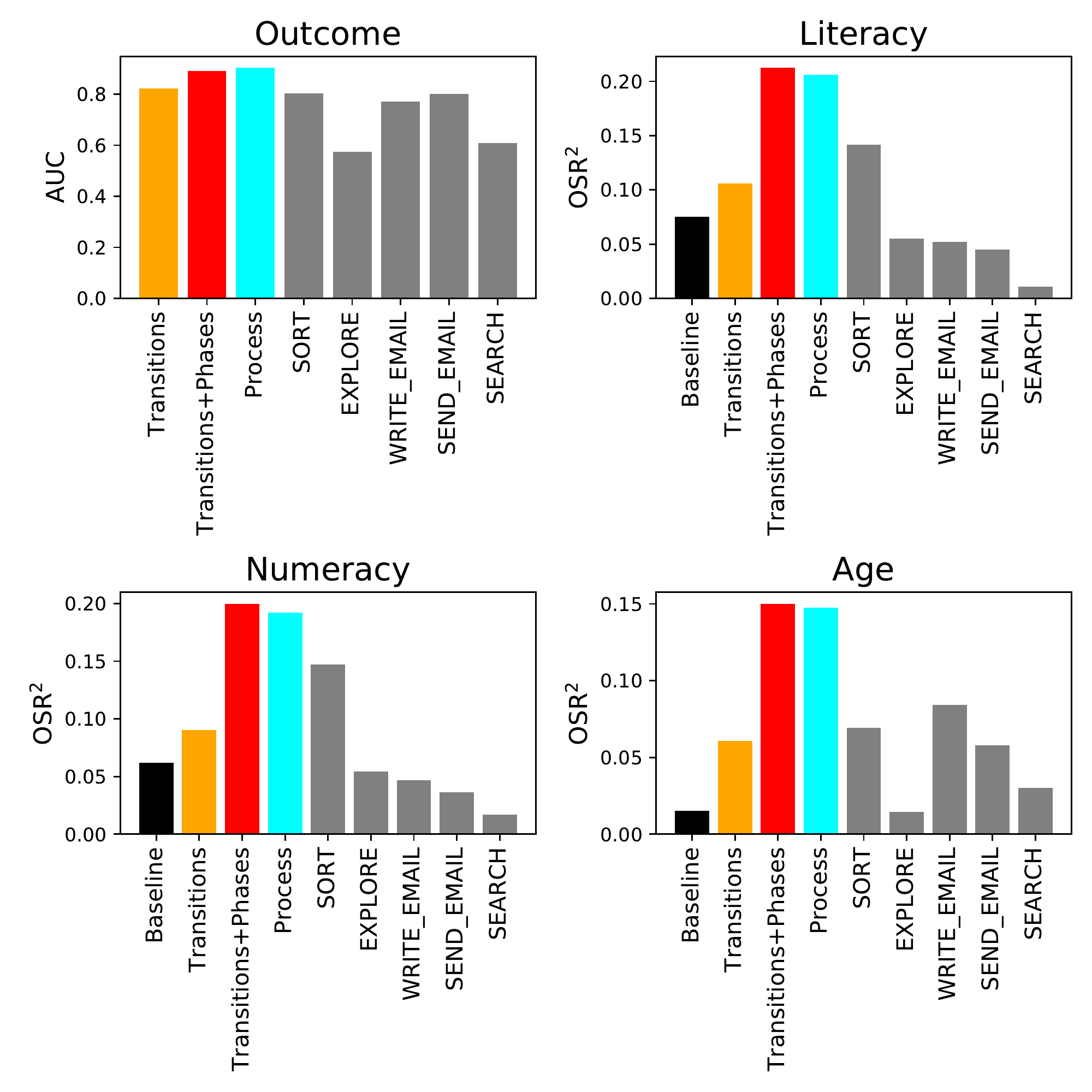}
	\caption{Out-of-sample $R^2$ under four different choices of features and the features from different subtasks of item U19a. The labels Baseline, Transitions, Transition+Subtasks and Process correspond to choice (i), (ii), (iii) and (iv) respectively. Other labels in grey correspond to the identified subtasks of U19a.} \label{fig:info_decomp}
\end{figure}

The prediction results are presented in Figure \ref{fig:info_decomp}.
A few observations can be made from the figure. 
First, the subtask sequence contains more information about respondents' literacy, numeracy scores and age than the binary final outcome as the $\text{OSR}^2$s corresponding to the subtask sequence are significantly higher than that of the baseline model.
Second, the amount of extra information provided by the detailed actions within subtasks depends on the variable of interest. For the binary final outcome, the subtask sequence carries only slightly less information than the original response process while for other variables, including detailed actions within subtasks can significantly improve the prediction performance.
Third, subtasks differ in the amount of information they can provide on a given variable of interest. 
Subtask SORT is most informative for predicting literacy and numeracy score, while subtask WRITE\_EMAIL produces the highest $\text{OSR}^2$ for predicting age.

	This information decomposition provides a structure of the response process and a map of dependence between the response process and other characteristics. 
	For example, if one wants to have an item that accurately reflects respondents' age, then including more email handling subtasks should be considered. The decomposition can also suggest real time intervention strategies. 
	If a respondent cannot not enter the subtask closely related to the final outcome after a long period of exploration or does not perform well in such subtasks, a hint may be provided to help the student stay on the right track.

\section{Concluding remarks}
Process data have been a recent focus in the psychometrics literature, due to the increased use of problem solving items in large scale testing as well as the recognition of potential value in educational assessment. Exploration of process data has been hindered by the lack of effective statistical and psychometric tools. In this paper, we develop a new approach to conducting exploratory analysis of process data. The new approach, using the concept of action predictability,  transforms long and noisy response processes into shorter and more interpretable subtask processes. Our case study using PIAAC 2012 shows that the subtask sequences can be used to visualize and identify respondents' problem-solving strategies. We also used the case study to show how the subtask sequences can be used to explore relationship between response processes and other variables of interest. 

The proposed approach allows a great deal of flexibility which entails possible refinement and improvement. Although we use an RNN-based model for the action prediction, other machine learning algorithms with a built-in sequential structure may serve as alternatives. In additional to the action sequence, process data also contain the sequence of action times. How to incorporate the action time stamps into the modeling and consequently the analysis should be of interest. We refer to \citep{wang2019modeling,de2019overview,lee2019how,zhang2018modeling} for recent works on analysis of response times in process data. 

\section{Acknowledgement}

The authors thank Educational Testing Service and the OECD for providing them with the PIAAC process data. 

\bibliographystyle{apalike}
\bibliography{subtask}

\appendix
\section{Structure of GRU}\label{sec:appendix:gru}
Let $\sigma$ denote the sigmoid function, i.e. $\sigma(x) = \frac{1}{1 + e^{-x}}$ and `$\star$' denote element-wise multiplication between vectors. The new hidden state $\boldsymbol{\theta}_{j+1}$, as shown by \eqref{eq:weighted_sum}, is a weighted sum of previous hidden state $\boldsymbol{\theta}_{j}$ and a candidate $\boldsymbol{\psi}_j$, while
$\boldsymbol{\kappa}_j$ and $\boldsymbol{r}_j$ is known as the update gate and reset gate respectively \citep{Cho2014phase}.




\begin{gather}
\boldsymbol{\kappa}_{j} = \sigma(U_1 \boldsymbol{\theta}_{j} + V_1 \boldsymbol{x}_j) , \label{eq:update_gate}\\
\boldsymbol{r}_{j} = \sigma(U_2\boldsymbol{\theta}_{j} + V_2 \boldsymbol{x}_j), \label{eq: reset_gate}\\
\boldsymbol{\psi}_{j}  = \tanh\left(U_3 (\boldsymbol{r}_j \star \boldsymbol{\theta}_j )+ V_3 \boldsymbol{x}_j\right),  \label{eq:psi}\\
\boldsymbol{\theta}_{j+1} = (\boldsymbol{1}-\boldsymbol{\kappa}_{j}) \star \boldsymbol{\theta}_{j} + \boldsymbol{\kappa}_{j} \star \boldsymbol{\psi}_{j}, \label{eq:weighted_sum}
\end{gather}

\section{Action Prediction Model Parameter Estimation}\label{sec:appendix_estimation}
The parameters in the action prediction model described in Section \ref{sec:action_pred} can be estimated by maximizing the log-likelihood function of $\bm \eta$ defined in \eqref{eq:generic loglik_peudo}. It is equivalent to minimizing the negative likelihood function $L(\bm \eta; \mathcal{S}) = -l(\bm \eta; \mathcal{S})$.
The estimator $\hat {\bm \eta} = \argmin_{\bm \eta} L\left(\bm\eta; \mathcal{S}\right)$ does not have a closed form. We adopt the stochastic gradient descent (SGD) algorithm \citep{Robbins1951SA} to approximate $\hat{\bm \eta}$ iteratively. In the SGD algorithm, $\bm \eta$ is initialized with some arbitrary value $\bm \eta^{(0)}$. Let $\bm \eta^{(g)}$ denote the parameter value after $g$ iterations, $g=1, 2, \ldots$. In iteration $g$, we randomly sample $i_g$ from $\{1, \ldots, N\}$ and update $\bm \eta$ according to  
\begin{equation}\label{eq:SGD}
\bm\eta^{(g)} = \bm\eta^{(g-1)} + \xi_g \nabla l\left(\bm\eta^{(g-1)}; \bm s^{(i_g)}\right),
\end{equation}
where $\xi_g$ is the step size of the update and $\nabla l \left(\bm\eta; \bm s^{(i)}\right)$ denote the gradient of $l(\bm \eta; \bm s^{(i)})$. Traditionally, the step size $\xi_g$ is often a predetermined decaying sequence. Several data-driven methods such as AdaGrad \citep{Duchi2011adaptive} and RmsProp \citep{Tieleman2012rmsprop} have been proposed recently. These methods compute $\xi_g$ from $\bm\eta^{(1)}, \ldots, \bm\eta^{(g-1)}$ and often leads to faster convergence in practice. We use RmsProp in synthetic and real data analysis.


Theoretically, the SGD algorithm should be run until the convergence of $\bm \eta$, that is, until the change of $\bm \eta$ between two consecutive iterations is below some threshold. 
However, as models involving neural network are often overparameterized, running the algorithm until convergence is often time-consuming and very likely leads to overfitting. 
To avoid these issues, we terminate the algorithm according to an early stopping rule. 
To apply this rule, the set of observed processes $\mathcal{S}$ is randomly split into a training and a validation set, denoted by $\mathcal{S}_{\text{train}}$ and $\mathcal{S}_{\text{valid}}$, respectively. 
We perform the SGD algorithm on the training set for a large enough number of iterations while monitoring the performance of the estimated model on the validation set. More specifically, in each iteration, $\bm s^{(i_g)}$ is sampled from $\mathcal{S}_{\text{train}}$ to update $\bm \eta$. Every several iterations, we evaluated $L(\bm \eta; \mathcal{S}_{\text{valid}})$ at the current value of $\bm \eta$. The set of parameter values with the lowest $L(\bm \eta; \mathcal{S}_{\text{valid}})$ is output as $\hat{\bm \eta}$. With this approach, we essentially terminates the algorithm before the model overfits.

The above algorithm is performed for a chosen $K$. To select an appropriate embedding dimension $K$, one can obtain $\hat{\bm \eta}$ for a range of $K$. The $K$ corresponding to the smallest validation loss $L(\hat{\bm\eta}; \mathcal{S}_{\text{valid}})$ is selected.

\section{Data generation models in simulation study}\label{sec:appendix:sim}
\subsection{Subtask sequence generation}
To generate the subtask sequence $\bm q$, we first uniformly sample the length of the subtask sequence $L$ from $\{3,4,5,6\}$. Given $L$, $\bm q = (q_1, \ldots, q_L)$ is generated from a Markov model with randomly generated starting distribution and state transition matrix. More specifically, we create a vector $\bm u = (u_i)_{i \in \mathcal{G}}$ and an $R \times R$ matrix $\bm U = \left(u_{ij}\right)_{i,j \in \mathcal{G}}$ by sampling their elements from a uniform distribution on $[0,1]$. The starting distribution $\bm \pi = \left(\pi_i\right)_{i \in \mathcal{G}}$ and the probability transition matrix $\bm P = \left(p_{ij}\right)_{i,j \in \mathcal{G}}$ are then computed by 
\begin{equation*}
\pi_i = u_i \Big/ \sum\limits_{j \in \mathcal{G}} u_j, \quad 
p_{ij} = \left\{
\begin{array}{ll}
0,  & j=i; \\
u_{ij} \Big/ \sum\limits_{k \in \mathcal{G}, k\neq i} u_{ik}, & j \neq i.
\end{array}
\right. 
\end{equation*}
The first element $q_1$ is generated according to $\bm \pi$. The remaining elements are generated iteratively based on the transition matrix.

\subsection{Subprocess generation}\label{sec:appendix:sim:subs}
For each subtask $g \in \mathcal{G}$, we uniformly sample without replacement from $\mathcal{A}$ a sequence of six special actions $\bm a^{(g)} = \left(a_1^{(g)}, \ldots, a_6^{(g)}\right)$, which serves as the standard solution to the subtask corresponding to subtask $g$. The actions not included in $\bm a^{(g)}$ are considered irrelevant for solving the subtask.
Given the subtask label $q_l = g$, we generate the $l$-th action subsequence $\tilde{\bm s}$ from a Markov model with the starting distribution $\bm \pi^{(g)} = \left(\pi_i^{(g)}\right)_{i \in \mathcal{A}}$ and the probability transition matrix $\bm P^{(g)} = \left(p_{ij}^{(g)}\right)_{i,j \in \mathcal{A}}$ specified below.
Let $\mathcal{A}^{(g)}$ be a set of the actions in the standard solution $\bm a^{(g)}$ except for the last one $a_6^{(g)}$. We generate actions in the subsequence until $a_6^{(g)}$ appears.
First, we generate a vector $\bm u^{(g)}  = \left(u_i^{(g)}\right)_{i \in \mathcal{A}}$ and an $M \times M$ matrix $\bm U^{(g)} = \left(u_{ij}^{(g)}\right)_{i,j \in \mathcal{A}}$ with elements independently generated from a uniform distribution on $[0,1]$. Then we generate a modified matrix $\widetilde{\bm U}^{(g)}  = \left(\tilde u_{ij}^{(g)}\right)_{i,j \in \mathcal{A}}$ from $\bm U^{(g)}$ to assign higher weights on $\mathcal{A}^{(g)}$ and the standard solution $\bm a^{(g)}$. Let $\psi$ be a random function on $\mathcal{A}$ such that 
$\psi\left(a_i^{(g)}\right) = a_{i+1}^{(g)}$ for $a_i^{(g)} \in \mathcal{A}^{(g)}$
and $\psi(a)$ is independently and uniformly sampled from $\mathcal{A}^{(g)}$ for $a \in \mathcal{A} \setminus \mathcal{A}^{(g)}$. 
By setting 
\begin{equation*}
\pi_i^{(g)} = u_i^{(g)} \Big/ \sum\limits_{j \in \mathcal{A}} u_j^{(g)}, \quad 
\tilde u_{ij}^{(g)} = \left\{
\begin{array}{ll}
100, & j = \psi\left(i\right) ;\\
u_{ij}^{(g)}, & j \neq \psi\left(i\right),
\end{array}
\right.  \quad 
p_{ij}^{(g)} = \tilde u_{ij}^{(g)} \Big/ \sum\limits_{k \in \mathcal{A}} \tilde u_{ik}^{(g)},
\end{equation*}
we get  the starting distribution $\bm \pi^{(g)}$ and the probability transition matrix $\bm P^{(g)} $ for generating action subsequences under subtask $g$. Hence, for each row indexed by action $i$, element in column $\psi(i)$ of $\bm P^{(g)} $ is a special action and also the most likely action in the next step. In particular, when $i$ is a special action in $\mathcal{A}^{(g)}$, $\psi(i)$ is the next special action in the standard solution $\bm a^{(g)}$.
\end{document}